# Examining Potential Usability and Health Beliefs Among Young Adults Using a Conversational Agent for HPV Vaccine Counseling

**Muhammad Amith, PhD[1*], Rebecca Lin[2*], Rachel Cunningham, MPH[3], Qiwei Luna Wu, MA[4], Lara S. Savas, PhD[1], Yang Gong, MD, PhD[1], Julie A. Boom, MD[3], Lu Tang, PhD[4], Cui Tao, PhD[1**],**

[1]The University of Texas Health Science Center at Houston, Houston, TX;
[2]Johns Hopkins University, Baltimore, MD;
[3]Texas Children's Hospital, Houston, TX; [4]Texas A&M University, College Station, TX.
* contributed equally to this work
** corresponding author, cui.tao@uth.tmc.edu

**Abstract**

*The human papillomavirus (HPV) vaccine is the most effective way to prevent HPV-related cancers. Integrating provider vaccine counseling is crucial to improving HPV vaccine completion rates. Automating the counseling experience through a conversational agent could help improve HPV vaccine coverage and reduce the burden of vaccine counseling for providers. In a previous study, we tested a simulated conversational agent that provided HPV vaccine counseling for parents using the Wizard of OZ protocol. In the current study, we assessed the conversational agent among young college adults (n=24), a population that may have missed the HPV vaccine during their adolescence when vaccination is recommended. We also administered surveys for system and voice usability, and for health beliefs concerning the HPV vaccine. Participants perceived the agent to have high usability that is slightly better or equivalent to other voice interactive interfaces, and there is some evidence that the agent impacted their beliefs concerning the harms, uncertainty, and risk denials for the HPV vaccine. Overall, this study demonstrates the potential for conversational agents to be an impactful tool for health promotion endeavors.*

**Introduction**

High-risk types of human papillomavirus (HPV) can lead to serious cancers in adults – including cervical, penile, head and neck, and several others. The most effective method to protect against HPV-related cancers is the HPV vaccine, which has been proven to protect against exposure to this virus[1]. Unfortunately, HPV vaccination rates continue to be lower than the targeted goal of 80%[2]. One predictive study mentioned that if at least 70% of the population were to receive the HPV vaccine, up to 6 million deaths could be prevented. One key tactic to increase the rate of HPV vaccination is patient-provider communication[3, 4]. In several studies, patient-provider communication has resulted in an effect on HPV vaccine uptake, as many patients and consumers prefer to discuss the HPV vaccine with their provider to make a decision[5-10]. Furthermore, patient-provider communication is a method endorsed by the President's Cancer Council to improve HPV vaccination rates[11].

The practice of patient-centered care has risen to prominence in recent years. Traditional medical care is physician-centered, in which physicians are seen to hold positions of authority over patients, but under the patient-centered approach, the patient's needs, feelings, concerns, and expectations are prioritized. This enables the physician to fully understand the patient's perspective to provide the best care possible. Effective communication between providers and patients is critical to facilitate exchange of information necessary for diagnosis and treatment while establishing mutual trust, involving both instrumental and affective behaviors.

Instrumental communication fulfills a patient's "need to know and understand," while affective communication fulfills a patient's "need to feel known and understood"[12]. More specifically, instrumental communication involves asking the patient questions about their symptoms, informing them of a diagnosis and treatment options, and discussing possible side effects. Meanwhile, affective communication involves being open and honest, expressing empathy, and addressing the patient by name—above all, viewing the patient as a person, rather than just a medical case. Only 7% of affective communication is verbal; non-verbal cues include tone of voice, eye contact, facial expressions, body language, and physical proximity[13]. Patients are extremely sensitive to these cues and may question their physician's genuineness if their verbal utterances seem inconsistent with their non-verbal behaviors (such as if the physician was withholding information about the severity and prognosis of the disease).

There are many positive effects of providing patient-centered care through instrumental and affective communication. These outcomes include increased patient satisfaction and understanding, which depend greatly on non-verbal affective behavior of physicians such as physical proximity[14]. On the other hand, dominant and controlling behaviors, such as frequent interruption to overshadow the patient in conversation, result in decreased patient satisfaction[15]. A physician may underestimate their patient's desire for information or fail to adequately communicate this information by using medical terms instead of everyday language, which may cause misunderstandings and lead to patient dissatisfaction and non-compliance. Additionally, patient-centered care is empowering; cancer patients who were given the opportunity to participate in decision-making about their treatment experienced higher quality of life and less anxiety[16].

Moreover, research has shown that the patient-centered approach produces optimal health outcomes, which is the ultimate aim of medical care. This may occur through direct or indirect pathways; for instance, clear communication would facilitate mutual understanding and trust, which could increase patient adherence and improve their condition[17]. Better-quality communication during history-taking and discussion of management plan is associated with symptom resolution, pain control, and physiological measures such as blood pressure and blood sugar level[18]. When patients perceive that their visit was patient-centered—especially if they found common ground with the physician for their management plan—they experienced better physical and emotional health[19]. Also, diagnostic tests and referrals two months after the visit were half as frequent if it was perceived to be patient-centered, which indicates increased efficiency of care.

Another important perspective to consider is that the implementation of patient-centered care addresses racial, ethnic, and socioeconomic disparities and improves access to health care for disadvantaged populations. These patient populations tend to ask fewer questions during visits and have lower health literacy, which often leads to worse outcomes[20-22]; for this reason, it is especially important to prioritize their perspective and help them fully understand their condition and treatment[23, 24]. Patients who lack health insurance and a consistent health care provider, or who have negative perceptions of their general health, also tend to rate provider-patient communication more unfavorably[25]. As these patients are most in need of effective communication with their physicians, open discussion should be encouraged to engage with them and help them access health services and community resources[17].

Despite the benefits of patient-centered communication, most physicians do not receive adequate training to develop these communication skills, which tend to deteriorate over time[26, 27]. In fact, complaints about clinical care typically arise due to poor communication, rather than technical competency issues. Common reasons for malpractice claims in the United States involve inadequate explanation of diagnosis or treatment and patients feeling ignored, devalued, or rushed[28]. Meanwhile, physicians tend to overestimate their communication skills: for instance, in a national survey conducted by the American Academy of Orthopedic Surgeons, 75% of orthopedic surgeons reported satisfactory communication with patients, but only 21% of patients reported satisfactory communication with their surgeons[29].

One potential concern of patient-centered care is that physicians "responding to every whim of the patient" may increase expenses to the health care system[19]. However, physicians with a more patient-centered communication style actually have fewer standardized diagnostic testing expenditures, as well as total standardized expenditures[30]. While these physicians tend to have longer visits with patients, there was no relationship between visit length and costs after adjusting for patient-centered communication scores. Nevertheless, longer visits are associated with preventive services, such as vaccination[28]. When recommending vaccines to patients, applying patient-centered communication is especially important to fully address their perceptions and concerns about vaccination, thereby promoting a shared understanding between physicians and patients that is essential for highly individualized clinical discourse.

The quality of a provider's recommendations for the HPV vaccine may have an impact on parents' decisions to vaccinate their children. For example, weak recommendations for the HPV vaccine are correlated with a lower likelihood of getting the HPV vaccine [31, 32]. A national survey also revealed hesitancy to recommend the vaccine and inconsistent recommendations among providers[33]. Some studies have found that one-third to one-half of eligible patients do not receive any recommendation for the HPV vaccine from their provider [3, 32]. Time constraints may also prevent providers from thoroughly discussing the HPV vaccine with their patients[34, 35].

The rising use of speech technology among consumers may open doors for patient-provider communication. Speech-based dialogue systems are software agents that carry discourse with a user involving several speech turns. We propose the possibility of automated discourse that could mimic the dialogue interaction between the patient and provider and respond consistently to each patient. This could inspire the patient to initiate discussion about the HPV vaccine, while offsetting some of the communication responsibilities of the provider by preparing the patient to ask specific questions and discuss their decision in detail when they see their provider. Some studies have also noted that young adults want to discuss their vaccination decisions with providers[36, 37], have misconceptions of the HPV vaccine and HPV[38], and are at high risk for HPV infection[39, 40]. Also, the younger demographic may have different informational needs compared to adults, such as the emphasis of sexual health practices to prevent HPV transmission. As such, an automated dialogue system may help in tailoring the discourse for individual users.

In a previous study[41], we performed an initial usability trial on using a conversational agent for HPV vaccine counseling among parents with at least one child under 18. The trial was conducted using the Wizard of OZ protocol that simulates natural language interface tools for agents or robots with a human operator ("wizard") in a remote area that is directing dialogue interaction of the machine[42]. The overall takeaway, despite some limitations on the speed of the "wizard" responses and the stoic speech, was receptiveness towards the ease of use and features of the agent, including interactivity of the dialogue and real-time question answering, useful information, and clarity of communication.

In this study, we applied the same procedure on a sample of college-aged adults at a Texas public university. HPV vaccination rates in Texas are much lower than in the rest of the United States (32.9% - 26.6% for boys and 39.7% for girls)[43]. This time, we collected more detailed data, including perceived beliefs of the HPV vaccine and responses from validated usability surveys. One aspect we were interested in exploring was how an agent would fare with similar systems that employed a voice interface, which would provide feedback critical to evaluate the effectiveness of the design of our HPV conversational agent. Furthermore, we wanted to know the potential impact of a conversational agent on vaccine uptake, which is a cue to action that may be predicted by health beliefs[44]. If there is some early evidence that a conversational agent may have a positive impact on health beliefs, this may warrant further study. Therefore, we propose the following questions:

- **RQ1:** *What is the overall usability of a simulated conversational agent compared to the usability of most interactive voice user interfaces?*

- **RQ2:** *Are there any correlations of the usability of the conversational agent with the various users' perceived beliefs towards the HPV vaccine?*

- **RQ3:** *Could a conversational agent for the HPV vaccine improve users' perceived beliefs?*

**Methods**

Participants were recruited from an undergraduate participant pool at Texas A&M University- College Station (TAMU) and chose to participate for an extra credit in a public speaking course[1]. Potential participants were young adults between the ages of 18 to 26. Out of 25 total participants, 11 were females and 14 were males, and the average age was 20 years. The majority were white (n=19) followed by Black (n=3), Hispanic (n=2) and Asian (n=1), respectively. 8 participants reported receiving the HPV vaccine, 8 reported not receiving the HPV vaccine, and 9 could not recall if they had received the vaccine. The study was conducted during the first week of April of 2019, utilizing the Wizard of OZ protocol[42] where the dialogue exchange is handled by a drone operator of the conversational agent (designated as "Beverly"). During the study, 1 participant experienced technical difficulties and was removed from analysis, resulting in a final sample size of 24.

Each participant was offered an explanation of the study (design, procedures, and risks) and was given time to ask questions; they were told that their participation was voluntary and that they could withdraw from the study at any time without negative consequence. If they agreed to participate, the study personnel obtained informed consent from them and the participant entered a designated observation area on the TAMU campus with a simulated conversational agent device (Wizard of OZ). The conversational agent, "Beverly," initiated a conversation with the user and about HPV and the HPV vaccine for 20-30 minutes. The drone operator utilized a script designed by study personnel to counsel on the HPV vaccine[41]. After the simulated counseling information session, each participant completed surveys to evaluate usability of the conversational agent and assess their health beliefs regarding the HPV vaccine.

---

[1] The University of Texas Health Science Center's Committee for the Protection of Human Subjects(HSC-SBMI-19-0102), and Texas A & M University Human Research Protection Program (IRB2019-0118M) approved this study.

The surveys were modified from three instruments: the System Usability Survey (SUS)[45, 46], the Speech User Interface Service Quality (SUISQ)[47, 48], and the Carolina HPV Immunization Attitude and Belief Scale (CHIAS)[49]. SUS is a validated industry standard to provide a "quick and dirty" scoring for usability, with a rating between 0-100 derived from 10 survey items. The SUS score can be interpreted by associating it with a letter grade (i.e. 80-89 = B, 70-79 = C, etc.) or an adjective rating[50, 51].

SUISQ is a 25-question survey developed specifically for interactive voice response applications based on four factors: User Goal Orientation (8 items), Customer Service Behaviors (8 items), Speech Characteristics (5 items), and Verbosity (4 items). User Goal Orientation describes the "system's efficiency, user trust, confidence in the system, and clarity of the speech interface." Customer Service Behaviors involve "friendliness and politeness of the system, its speaking pace, and its use of familiar terms." Speech Characteristics refer to "naturalness and enthusiasm of the system voice," and Verbosity is the "talkativeness and repetitiveness of the system[52]." Each item is on a Likert scale between 1 to 7 (1 for strongly disagree to 7 for strongly agree). The factor scores are the mean of all items in each category and an overall score is the mean of the factor scores (Verbosity mean is reversed).

CHIAS is a 16-item survey with four factors (Perceived Barriers, Harms, Effectiveness, and Uncertainty) that are aligned with the health belief model[44], a well-known health behavioral change model that has a long history in vaccine uptake research. Survey items for Perceived Barriers relate to "barriers to HPV vaccination including cost and access to a healthcare provider"[49]. Perceived Harms pertains to "perceived potential harms from the vaccine including health problems"[49]. Perceived Effectiveness and Uncertainty survey items describe subjective notions of the "effectiveness of HPV vaccine in protecting against genital warts and cervical cancer"[49] and "not having enough information about the HPV vaccine and perception of community vaccination norms"[49], respectively. Each survey item is on a Likert scale (1 for strongly disagree to 4 strongly agree). Lower ratings represent positive attitudes towards the HPV vaccine. This survey has demonstrated stability of the factors to describe HPV vaccination attitudes over time. With the exception of the Perceived Effectiveness factor, the other three factors have been attributed to parent HPV vaccination intention and predicted HPV vaccination utilization of parents[53]. CHIAS is adaptable for young adults and has been shown to have similar validity to the parent version[54, 55]. This version targeted for young adults includes one additional factor called Risk Denial[55], a factor relating to low perception of risk of the HPV infection. Two recent studies utilized this variation of CHIAS on samples of a young population under 25. Kamimura and colleagues utilized the Risk Denial and Benefit factors of the survey to measure differences between Vietnamese and United States college students[56], while Hanson and associates focused on and measured the Perceived Effectiveness, Uncertainty, and Harms factors of the CHIAS model for their survey study[57].

We utilized the same software system that we developed to provide speech dialogue interaction with a natural language (NL) interface for the user[41]. The software was developed using iOS SDK for the tablet and Java for the desktop application to remotely interact with the user through the tablet's NL speech interface. We also employed the same script in our previous trial, but we modified the language to reflect the target demographic. For example, a scripted statement like "[i]f **_your child_** is vaccinated with the HPV vaccine it will protect against various HPV viruses which causes many precancerous and cancerous lesions in males and females" became "if **_you are_** vaccinated with the HPV vaccine..."

**Results**

We collected survey responses from patients using the aforementioned instruments (SUS, SUISQ, and CHIAS). For four questions of the CHIAS survey, we reversed the scale to match the direction of the other CHIAS variables so that higher scores were indicative of lower receptiveness for the HPV vaccine. For the SUS survey, we omitted the question "I think I would like to use the system frequently," because participants were unlikely to use the system again. According to Lewis & Sauro analyses, removal of this question would not have a major impact on the statistical validity of the SUS final score[58]. To compensate for the omitted question, we adjusted the SUS calculation[58] because we do not foresee this agent being used every day as a consumer tool. Rather, we envisioned the agent to be situated in a clinical environment while the patient is waiting for their health care provider. Statistical analysis was completed through IBM SPSS v25.

**RQ1**: For the primary question regarding overall usability, we computed a one sample t-test against a reported average SUS score for voice interfaces, μ = 72 (n = 233)[51]. The analysis produced a t-statistic of t(23)=1.627, p=0.059. Overall, there is 94% confidence that the system has a score above the industry mean of 72. This would reveal that the conversational agent had an on-par or a slightly above average SUS score with industry-based interactive voice interface systems.

We also looked at the differences between users by vaccination status (See Figure 1). The SUS overall score was higher among those that never had the vaccine (μ = 80) compared to those that did have the HPV vaccine (μ = 77) and those did not know if they had the HPV vaccine (μ = 74). We also looked at the attributes of individuals in relation to the SUS score to ascertain any association with the SUS Score. A Fisher Exact Test (p = 0.54, 12.42) indicated a statistically marginally significant relationship between vaccination status and SUS score. None of the other attributes such as school classification, parents' income, etc. indicated any significance with the SUS Score.

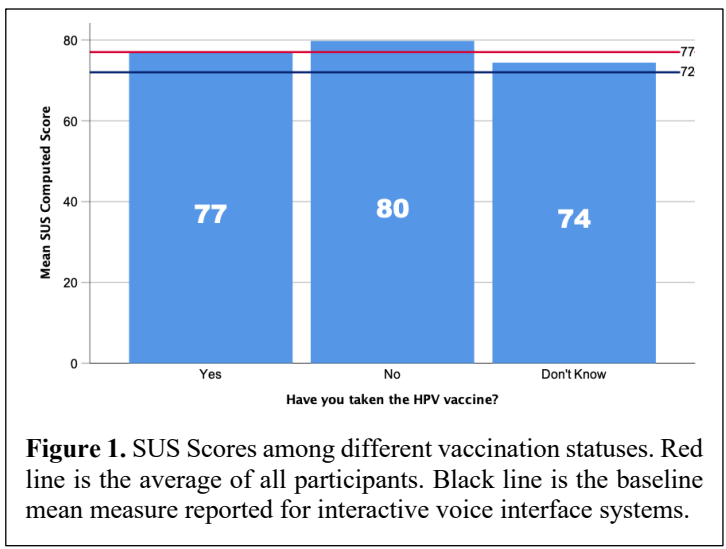

**Figure 1.** SUS Scores among different vaccination statuses. Red line is the average of all participants. Black line is the baseline mean measure reported for interactive voice interface systems.

While the SUS survey has a history of being a valid measurement tool for quick usability scoring, the SUISQ survey has yet to be as standardized as the SUS. We administered the SUISQ survey (See Table 1) in conjunction with SUS, as recommended by Lewis[59]. We looked at the correlation between SUS and SUISQ to ascertain any significant relationship. For the most part, there appears to be a positive correlation between SUS score and SUISQ score that is moderately strong and statically significant (r=0.486, p=0.016). For the User Goal Orientation and Verbosity factors of SUISQ, there also appears to be a strong, positive relationship with the SUS score (r=0.508, p=0.005 and r=0.627, p=0.000, respectively). There is also a moderately positive correlation between Customer Service Behavior score and SUS score (r=0.438, p=0.014).

**RQ2**: To answer RQ2, we investigated any correlations for usability with the participants' beliefs of the HPV vaccine. In comparing those who reported to not have received the HPV vaccine with those who reported to have received the HPV vaccine, we calculated the Spearman rank correlation. Among those that reported to not have received the HPV vaccine, there was a strong inverse relationship (i.e., a lower CHIAS rating indicates a greater propensity for the HPV vaccine) between SUISQ and the Perceived Effectiveness construct, which was statistically significant (rs= -.711, p=0.037). There was also a strong inverse relationship between the Verbosity factor of SUISQ and the

**Table 1**. SUISQ Factor means and overall score.

| SUISQ Factor | Mean (SD), n=24 |
|---|---|
| User Goal Orientation | 4.69 (0.96) |
| Customer Service Behavior | 6.09 (0.60) |
| Speech Characteristics | 3.10 (1.29) |
| Verbosity (reversed) | 3.26 (1.03) |
| **SUISQ Overall Score** | **4.29 (0.75)** |

Perceived Uncertainty construct for the CHIAS, which was statistically significant (rs= -0.874, p=0.05). Among those that reported to have received the HPV vaccine, there was a strong inverse relationship between the overall SUISQ score and the Perceived Barriers and Perceived Effectiveness constructs, which were statistically significant (rs= -0.639, p=0.044 and rs= -0.655, p=0.039, respectively). In addition, there was a strong inverse relationship between the Verbosity factor of SUISQ and the Perceived Effectiveness construct, which was statistically significant (rs= -0.764, p=0.014).

**RQ3**: Lastly, we investigated if there was any improvement in the users' perception of the HPV vaccine compared to samples from published studies conducted by Kamimura et al.[56] and Hanson et al.[57]. We re-scaled our collected CHIAS data to match the scales for Kamimura's and Hanson's CHIAS data – a 5-point and 11-point scale, respectively. Then we computed a one-sample t-test with the means of the CHIAS factors using the CHIAS factors of the aforementioned studies. Table 2 summarizes our results where the green signifies statistical significance, gray signifies marginal statistical significance and red signifies no statistical significance.

There appears to be better beliefs and perceptions with Risk Denial ($\mu$ =1.41, p=0.00), Perceived Harms ($\mu$ =1.92, p=0.00) and Perceived Uncertainty ($\mu$ =2.13, p=0.00) among the individuals who interacted with our conversational agent. There was some improvement if accounting for the marginally statistical significance with Perceived Barriers ($\mu$ =1.94, p=0.09). There was no evidence of improvement with Perceived Effectiveness despite a better rating ($\mu$ =3.82, p=0.26).

Table 2. Comparison of previous studies' means of CHIAS constructs between agent and published studies. Lower values for CHIAS indicate positive attitudes in favor of the HPV vaccine. Green rows indicate statistical significance, gray rows represent marginally statistical significance, and red represents no statistical significance. Adjusted p-value calculated with Holm-Bonferroni using alpha level of .05. $\pm$Kamimura, et, al.,+Hanson, et, al.

| CHIAS Construct | Agent + CHIAS | CHIAS | p-value | adjusted p-value |
|---|---|---|---|---|
| Perceived Barriers | 1.94 | 2.33 (n=437)$^\pm$ | 0.09 | .0125 |
| Risk Denial | 1.41 | 2.00 (n=437) | 0.00 | .01 |
| Perceived Harms | 1.92 | 3.5 (n=108)$^+$ | 0.00 | .01 |
| Perceived Effectiveness | 3.82 | 4.3 (n=108) | 0.26 | .02 |
| Perceived Uncertainty | 2.13 | 4.8 (n=108) | 0.00 | .01 |

**Discussion**

This second attempt at the Wizard of OZ experiment was performed on young adults who are also potential targets for the HPV vaccine, considering the recently expanded approval of the vaccine for adults up to 45 years of age[60]. In this study, we administered a more detailed usability survey involving the System Usability Scale (SUS) and Speech User Interface Service Quality (SUISQ). SUS is a general usability measurement scale based on 10 Likert scaled questions that has proven reliability. While SUISQ is a more robust usability survey for voice user interfaces, it has yet to achieve the reliability of the SUS. Together they may provide further validation support for SUISQ[59]. We also administered a health belief model survey, the Carolina HPV Immunization Attitudes and Beliefs Scale (CHIAS), that was tailored for young adults[55] to see if there were any correlations between the health belief constructs and usability variables. We compared the means for each of the constructs with those from previous studies that used this modified survey for a younger population[56, 57].

**RQ1:** The overall usability (i.e. satisfaction, effectiveness, and efficiency), measured through the SUS score, indicated the conversational agent had average to slightly better scores than many of the interactive voice interface systems (a SUS score of 77 versus 72; 94% confidence). Additionally, there was a marginally statistically significant relationship between SUS score and HPV vaccination status. There were slight variations of the SUS score among the different users in comparing those that had received the HPV vaccine (SUS score of 77) with those that did not know if they had received the vaccine (SUS score of 74). The individuals that never received the HPV vaccine rated the system high at 80. This score indicates that this tool could be effective for individual users who had never received the HPV vaccine and could perhaps inform and encourage them better than other paper-based methods. With a small sample size, we can conclude that there is preliminary evidence that our conversational agent (while simulated) has a strong usability for people who never received the HPV vaccine. In the future, we can further analyze why those that didn't know if they received the HPV vaccine scored lower and perhaps look into methods to improve SUS scores among this group of individuals.

**RQ2:** Lower values for the health belief ratings indicated a propensity for the HPV vaccine. Therefore, we sought any statistical significance showing an inverse relationship between the CHIAS constructs and usability factors. No significant relationship exists between SUS and the CHIAS constructs, but there are some statistically significant relationships between the SUISQ factors and the CHIAS constructs. Among the individuals who never received the HPV vaccine, there was a strong inverse relationship between SUISQ Overall Score and Perceived Effectiveness of the HPV vaccine. This would indicate that unvaccinated participants who liked the conversational agent are more likely to believe that the HPV vaccine is effective and consider getting vaccinated. Within that same group, the

Verbosity factor — measuring how repetitive and talkative the agent is — had a significant inverse relationship with the Perceived Uncertainty construct. For background, the Verbosity score was reversed for analysis, so a high Verbosity score indicated low repetitiveness and talkativeness of the agent. Perceived Uncertainty represents sufficiency of information concerning the HPV vaccine. What we gather from our preliminary results for the relationship between Verbosity and Perceived Uncertainty is that the content and information related to the vaccine and HPV-related diseases that was spoken by the system was not deemed meaningless and was adequate for the user.

Of interest, the individuals who reported to have received the HPV vaccine also shared the same relationship pairing — Verbosity ≈ Perceived Uncertainty, and Overall SUISQ ≈ Perceived Effectiveness. Also, SUISQ had a significant relationship with the Perceived Uncertainty and Perceived Barriers constructs. The Perceived Barriers rating reveals any external challenges that could impede obtaining the HPV vaccine. Since individuals reported having the HPV vaccine, it is clear they overcame any barriers that could have prevented them from obtaining the HPV vaccine. The various relationships with several variables are more likely due to them already having the HPV vaccine, since the SUISQ overall score was associated with three of the five CHIAS constructs.

**RQ3:** When we analyzed the Carolina HPV Immunization Attitudes and Beliefs Scale (CHIAS) user data with the CHIAS mean results from other published studies that used the same survey[56, 57], there was a statistical difference (i.e., improvement) with various means of the constructs from the CHIAS. Aside from the Perceived Effectiveness construct, there was also a marginally significant statistical difference with the Perceived Barriers construct. In absence of a true control group, there is some evidence of impact on health beliefs and perceptions of the HPV vaccine as a result of using a conversational agent for the HPV vaccine.

**Limitations**: This study has several important limitations. The most significant limitation is that the study lacked a control group. Because we needed to have as many participants as possible to provide stronger analysis of the results, we had all of the participants interact with the conversational agent and used the means data for some of the constructs to compare with existing studies. Also, with the lack of resources and time (and the nature of human computer-interaction studies), the sample size was relatively small and is not generalizable to a larger population. With a larger sample size, some of the marginally significant statistical findings could have had more validity. Among those that reported not knowing whether they received the HPV vaccine, no significant inverse relationship with any of the variables existed, so we need to investigate why this was the case. It might be possible that they have low health literacy, but we cannot be certain unless we measure their health literacy skills. It may also be due to their lack of interest in their personal health, in which case a tailored dialogue intervention may be needed for this group. Another limitation is that the age group recruited through this study (young adults) may differs from another intended demographic who will use our conversational agent, namely parents of pediatric patients. Although parents were receptive to the usability of the conversational agent in a previous study[41], we have not yet gathered data on their beliefs towards the HPV vaccine, which may not necessarily be the same as those of college-aged adults.

**Conclusion**

Overall, based on the initial analysis, we conclude that our initial design for a conversational agent for the HPV vaccine is robust and likely to be consistent with industry standards. Future assessments with a larger, more diverse population would allow us to further validate the usability and likeability of our conversational agent. Also, the usability may have some correlation with the health beliefs of its users as it relates to the HPV vaccine. Lastly, based on comparisons with previous studies assessing health beliefs among young adults, the conversational agent could potentially impact the health beliefs of the users as it pertains to the HPV vaccine. This study provides initial evidence on the validity of using a conversational agent to improve vaccination rates. While these results are preliminary at best and lack a large sample size, there is some promise that deploying this type of tool within a health care environment could be feasible, especially since conversational agents promote patient-centered communication, which is important to achieve better health outcomes. Also, the relatively good usability rating which encompasses overall satisfaction could be analogous to patient satisfaction, which is an important outcome of patient-centered communication that we discussed earlier. In addition, the implementation of conversational agents in a health setting could possibly address some of the communication challenges rooted in demographic disparities (language barriers, low health literacy, etc.), and offload some of the communication burdens that impede on providers' time. However, additional research is needed to further validate our findings. This work could potentially be extended to other vaccine needs and health topics as well.

**Future Direction**: In this study and the last, we utilized the Wizard of OZ protocol to simulate a conversational agent for the HPV vaccine as a pilot study. Our next endeavor is to provide a fully automated agent to provide discourse on the HPV vaccine with health consumers. Currently, we have developed some of the knowledge bases[61-64] and software components that can provide this interaction[65, 66], and we plan on live testing them sometime in the near future with both parents and young adults. Overall, this experience with the simulated system provided us with knowledge for future development and planning for an autonomous conversational agent that can counsel patients on receiving the HPV vaccine, with the potential to be extended towards other health promotion topics.

**Acknowledgements**

Research was supported by the UTHealth Innovation for Cancer Prevention Research Training Program (Cancer Prevention and Research Institute of Texas grant # RP160015), the National Library of Medicine of the National Institutes of Health under Award Number R01LM011829, and the National Institute of Allergy and Infectious Diseases of the National Institutes of Health under Award Number R01AI130460.